\documentstyle[aps,pra,epsf]{revtex}
\begin{document}
\draft
\newcommand{\pp}[1]{\phantom{#1}}
\newcommand{\be}{\begin{eqnarray}}
\newcommand{\ee}{\end{eqnarray}}

\title{
Einstein-Podolsky-Rosen-Bohm experiment
with relativistic massive particles
}
\author{Marek Czachor \cite{*}}
\address{
Katedra Fizyki Teoretycznej i Metod Matematycznych\\
 Politechnika Gda\'{n}ska,
ul. Narutowicza 11/12, 80-952 Gda\'{n}sk, Poland
}
\maketitle
\begin{abstract}
Two aspects of the relativistic version of the 
Einstein-Podolsky-Rosen-Bohm (EPRB) experiment with massive particles
are discussed: (a) a possibility of using the experiment as an
implicit test of a relativistic center-of-mass concept, and 
(b) influence of the relativistic effects on degree of
violation of the Bell inequality. 
The nonrelativistic singlet state average 
$\langle \psi|{\vec a}\cdot\vec \sigma\otimes  {\vec
b}\cdot\vec \sigma|\psi\rangle =-\vec a\cdot\vec b $ 
is relativistically
generalized by defining spin {\it via\/} the relativistic
center-of-mass operator. The corresponding EPRB average contains
relativistic corrections which are stronger in magnitude 
than standard relativistic phenomena such as the time delay, and can be
measured in Einstein-Podolsky-Rosen-Bohm-type experiments with
relativistic massive spin-1/2 particles. 
The degree of violation of the Bell
inequality is shown to depend on the velocity of the pair of
spin-1/2 particles with respect to laboratory. 
Experimental confirmation of 
the relativistic formula would indicate that 
for relativistic nonzero-spin particles centers of mass and
charge do not coincide. The result may have implications for quantum
cryptography based on massive particles. 
\end{abstract}
\vskip1pc

\section{Introduction}

Contemporary applications of the 
Einstein, Podolsky and Rosen (EPR) correlations
\cite{EPR,Bohm} and the Bell inequality \cite{Bell,Home} range from
purely philosophical problems to 
quantum cryptography, computation 
and teleportation.

In the cryptographic scheme proposed by Ekert \cite{Ekert} Alice
and Bob test for eavesdropping by measuring the average 
of the ``Bell observable"
\begin{eqnarray}
c(\bbox a,\bbox a',\bbox b,\bbox b')&=&
\langle \psi|\hat {\bbox a}\otimes  \hat {\bbox
b}|\psi\rangle
+
\langle \psi|\hat {\bbox a}\otimes  \hat {\bbox
b'}|\psi\rangle\nonumber\\
&\pp =&
+
\langle \psi|\hat {\bbox a'}\otimes  \hat {\bbox
b}|\psi\rangle
-
\langle \psi|\hat {\bbox a'}\otimes  \hat {\bbox
b'}|\psi\rangle\label{Bell}
\end{eqnarray}
where $\hat {\bbox a}$ etc. are ``yes--no" observables (say, signs of
spin for electrons, or planes of polarization for photons).
Quantum mechanics predicts that for some choices of such
observables one can obtain
$|c(\bbox a,\bbox a',\bbox b,\bbox b')|=2\sqrt{2}$. In an ideal
situation  a result of the form 
$|c(\bbox a,\bbox a',\bbox b,\bbox b')|<2\sqrt{2}$ indicates that
at least some pairs of particles were not prepared in the singlet
state and this indicates an eavesdropper. 

Practical applicability of quantum cryptographic protocols
crucially depends on detector efficiencies. In typical Bell-type
photon pair 
experiments the efficiencies were smaller than $20\%$. 
The advent of solid state
photodiodes provides efficiencies of detection
which are  much higher 
\cite{Kwiat} but still far from ideal.

An almost ideal experimental scheme has been recently discussed
by Fry {\it et al.\/} \cite{Fry} who propose 
to replace photons with massive particles (pairs of $^{199}$Hg atoms). 
Detection efficiency
is then at least $95\%$ and can be pushed to more than
$99\%$. An obvious drawback of such a communication channel is
that it is slow. To make it faster one might be tempted to use
relativistic velocities. 

It will be shown below that for high velocities one may expect a
surprising effect: The amount of violation of the Bell
inequality may decrease with growing velocity of the spin-1/2
particles. 
Alice and Bob must therefore additionally know the velocity distribution
of the particle beam. Otherwise they may be confused and
``detect an eavesdropper" even though the particles remain in a
pure zero-helicity singlet state.
The effect is related to the old problem described already in
1930 by Schr\"odinger \cite{ES}.
As is widely known
E.~Schr\"odinger examined
the behavior of the coordinate operator $\bbox x$ 
associated
with Dirac's equation and discovered the oscillatory motion 
he called the {\it Zitterbewegung\/}. The {\it Zitterbewegung\/}
takes place with respect to the {\it center-of-mass\/} position
operator $\bbox x_A$ and this is the operator which should be
used to define a physically meaningful spin operator.
The situation is not typical only of the Dirac equation and is
not associated with the presence of negative energy solutions as
one is sometimes led to believe.
The so-called new Dirac equation generalized by
Mukunda {\it et al.\/}  \cite{Mukunda} admits only
positive-energy solutions but the {\it Zitterbewegung\/} is
present and the associated center-of-mass operator is
algebraically identical to this implied by Schr\"odinger's
analysis of the Dirac equation \cite{BZ}. The analysis
presented in \cite{Mukunda} shows clearly that in order to
obtain a physically consistent model of an extended hadron one
has to proceed in the way identical to the one chosen in this
paper: First define the center-of-mass operator $\bbox Q$, then introduce 
the angular momentum $\bbox L=\bbox Q\times\bbox P$, and finally define
spin by $\bbox S=\bbox J -\bbox L$. 

In what follows I  use a group representation formulation,
elements of which can be found in
the 1965 papers by Fleming \cite{Fleming1}. The group theoretic
approach has the advantage of being applicable to any physical
system whose symmetry group is the Poincar\'e group, or whose
symmetry group contains the Poincar\'e group as a subgroup. 

\section{Relativistic spin operators}

Let us begin with generators of the unitary, infinite
dimensional irreducible representation of the Poincar\'e group
corresponding to a nonzero mass $m$ and spin $j$. Their standard
form is \cite{Ohnuki}
\begin{eqnarray}
\bbox J &=& 
\frac{\hbar}{i}\bbox p\times \frac{\partial}{\partial \bbox p}
+{\bbox s},\\
\bbox K &=& 
\pm\Bigl(
|p_0|\frac{\hbar}{i}\frac{\partial}{\partial \bbox p}
- \frac{\bbox p\times{\bbox s}}{mc+|p_0|}\Bigr),\label{46}\\
\bbox P&=& \bbox p,\\
P_0&=&p_0=\pm \sqrt{\bbox p^2 + m^2c^2}.
\end{eqnarray}
Here {\bbox s} denotes finite dimensional angular 
momentum matrices corresponding to
 the $(2j+1)$-dimensional representation 
$D^j$ of the rotation group. Similar forms are obtained if one
uses the hadronic representation introduced in \cite{Mukunda}.

The Poincar\'e group has two Casimir operators: The squared mass
and the square of the Pauli-Lubanski vector $W^\mu$. The latter
operator written in the above representation is
\begin{eqnarray}
W^\mu&=&(W^0,\bbox W)=(\bbox P\cdot\bbox J,P_0\bbox J -\bbox
P\times\bbox K) \\
&=&
\bigl(\bbox p\cdot \bbox s, p_0(\bbox n\cdot \bbox s)\bbox n
\pm mc\, \bbox s_\perp\bigr),
\end{eqnarray}
where $\bbox n$ is the unit vector pointing in the momentum
direction and 
$$
\bbox s_\perp=\bbox s-(\bbox n\cdot \bbox s)\bbox
n.
$$ 
The center-of-mass position operator which generalizes to any
representation the operator $\bbox x_A$ of Schr\"odinger is
\begin{eqnarray}
{\bbox Q} &=&-\frac{1}{2}\Bigl(P_0^{-1}{\bbox K} +{\bf
K}P_0^{-1}\Bigr) \label{Q}\\
&=&i\hbar\frac{\partial}{\partial \bbox p} 
-i\hbar\frac{\bbox p}{2p_0^2}
+ \frac{\bbox p\times{\bbox s}}{|p_0|(mc+|p_0|)}.\label{conn}
\end{eqnarray}
This operator extends naturally also to massless fields
and can be shown to be uniquely (up to subtleties with domains
of unbounded operators) derived from symmetry considerations
in the case of the Maxwell field \cite{JJ,Mourad}. In the Maxwell field
case, the formula (\ref{conn}) can be regarded as 
defining a connection on a light cone. A parallel transport with
respect to this connection can be shown to generate a Berry
phase \cite{IBB,Pati}.

Orbital angular momentum and spin corresponding to $\bbox Q$ 
were given by Pryce and Fleming 
\cite{Fleming1,Pryce}
\begin{eqnarray}
\bbox L &=& \bbox Q\times\bbox P=
\frac{\hbar}{i}\bbox p\times \frac{\partial}{\partial \bbox p}
+\frac{|p_0|-mc}{|p_0|}\Bigl(
{\bbox s} -(\bbox n\cdot{\bbox s})\bbox n\Bigr), \nonumber\\
{\bbox S} &=& \bbox J - \bbox L = \frac{mc}{|p_0|}{\bbox s} +
\Bigl(1-\frac{mc}{|p_0|}\Bigr)(\bbox n\cdot{\bbox s})\bbox n
\nonumber\\
&=&
\sqrt{1-\beta^2}{\bbox s}_\perp+(\bbox n\cdot{\bbox s})\bbox n
=
\bbox W/p_0.\label{s-w}
\end{eqnarray}
$\bbox \beta=\bbox n\,|\bbox v|/c$, where $\bbox v=c^2\bbox p/p_0$ 
is a velocity of the particle.
Eq.~(\ref{s-w}) shows that relativistic spin is closely related
to the Pauli-Lubanski vector. 
Projection of spin in a direction given by the unit vector
$\bbox a$ commutes with the Hamiltonian $P_0$ and equals
\begin{eqnarray}
\bbox a\cdot{\bbox S}&=&
\Bigl[\frac{mc}{|p_0|}{\bbox a} +
\Bigl(1-\frac{mc}{|p_0|}\Bigr)(\bbox n\cdot{\bbox a})\bbox n
\Bigr]\cdot{\bbox s}\\
&=&
\Bigl[\sqrt{1-\beta^2}\bbox a_\perp +\bbox
a_\parallel\Bigr]\cdot \bbox s=:
\bbox \alpha(\bbox a,\bbox p)\cdot{\bbox s}.
\label{9}
\end{eqnarray}
The latter equality 
defines the vector $\bbox \alpha(\bbox a,\bbox p)$ whose length
is 
\begin{eqnarray}
|\bbox \alpha(\bbox a,\bbox p)|=
\frac{\sqrt{(\bbox p\cdot\bbox a)^2 +m^2c^2}}
{|p_0|}=\sqrt{1+(\bbox \beta\cdot\bbox a)^2-\beta^2}.\nonumber
\end{eqnarray}
The eigenvalues of $\bbox a\cdot{\bbox S}$ are therefore
\begin{eqnarray}
\lambda_{a}=j_3\hbar
\sqrt{1+(\bbox \beta\cdot\bbox a)^2-\beta^2}
\label{aS}
\end{eqnarray}
where 
$j_3=-j,\dots,+j$.
The eigenvalues of the Pauli-Lubanski vector projections are
therefore $\omega_a=p_0\lambda_a$.
In the infinite momentum/massless limit the eigenvalues of 
the relativistic spin
in a direction perpendicular to $\bbox p$ vanish, which can be
regarded as a consequence of the Lorentz flattenning of the
moving particle (in these limits $\bbox S=(\bbox n\cdot\bbox
s)\bbox n$). 
Projection of spin on the momentum direction is
equal to the helicity, i.e.
$\bbox p\cdot\bbox S=\bbox p\cdot\bbox s$ for any $\bbox
p$, and 
 $\bbox S=\bbox s$ in the rest frame ($\bbox p=0$). 
Bacry \cite{Bacry} observed that a nonrelativistic limit of
$\bbox Q$ leads to a correct form of the spin-orbit
interaction in the Pauli equation if one uses potentials
$V(\bbox Q)$ instead of $V(\bbox x)$ \cite{Kaiser}; 
an analogous effect was
described in \cite{BB1} where the internal angular momentum of
the {\it Zitterbewegung\/} leads to spin with the correct $g=2$
factor. An algebraic curiosity is the fact that the
components of $\bbox S$ satisfy an algebra which is $so(3)$ in
the rest frame and formally
contracts to the Euclidean $e(2)$
in the infinite momentum/massless limit, and thus
provides an interesting alternative 
explanation of the privileged role
played by the Euclidean group in the theory of massless fields
\cite{ja,Kim}.

In spite of all these facts suggestng
that both $\bbox Q$ and $\bbox S$ are
natural candidates for physical observables no experimental tests
distinguishing them from other definitions of position and spin
have been proposed so far. Obviously, it is not easy to test
directly $\bbox Q$ which, representing the center of mass, may be
expected to couple to the gravitational field. The spin
operator, on the other hand, is responsible for the magnetic
moment and should couple to the electromagnetic field which is
much stronger.

Consider now a Stern-Gerlach-type measurement involving spin-1/2
relativistic particles and assume that $\bbox S$ is 
the physical internal angular
momentum which is measured in this experiment \cite{some}. 
Assume also that we have  two spin-1/2 particles in a singlet state (total
helicity equals zero) and propagating in the same direction with
identical momenta $\bbox p$ (to be more precise one should take wave
packets in momentum space, but for simplicity assume that they
are sufficiently well localized around momenta $\bbox p$, so
that we can approximate them by plane waves):
\begin{eqnarray}
|\psi\rangle = \frac{1}{\sqrt{2}}\Bigl(
|+\frac{1}{2},\bbox p\rangle|-\frac{1}{2},\bbox p\rangle 
-
|-\frac{1}{2},\bbox p\rangle|+\frac{1}{2},\bbox p\rangle
\Bigr). \label{st}
\end{eqnarray}
The kets $|\pm\frac{1}{2},\bbox p\rangle$ form the {\it helicity\/} basis.
Consider the binary operators
$\hat {\bbox a}=\bbox a\cdot\bbox S/|\lambda_a|$,
$\hat {\bbox b}=\bbox b\cdot\bbox S/|\lambda_b|$.
Their eigenvalues are $\pm 1$. The relativistic corrections that
arise are those resulting from the modification of the spin
direction as ``seen" by a measuring device. 
The average of the relativistic EPR-Bohm-Bell operator is 
\begin{eqnarray}
{}&{}&
\langle \psi|\hat {\bbox a}\otimes \hat {\bbox
b}|\psi\rangle =\nonumber\\
&{}&
- \frac{
\bbox a\cdot\bbox b- \beta^2 \bbox a_\perp\cdot\bbox b_\perp}
{\sqrt{1+\beta^2\bigl[(\bbox n\cdot\bbox a)^2 - 1\bigr]}
\sqrt{1+\beta^2\bigl[(\bbox n\cdot\bbox b)^2 - 1\bigr]}}\label{ab}
\end{eqnarray}
There are several interesting particular cases of the formula
(\ref{ab}). 
First, if $\bbox a=\bbox a_\perp$, 
$\bbox b=\bbox b_\perp$
then 
\begin{eqnarray}
\langle \psi|\hat {\bbox a}\otimes \hat {\bbox
b}|\psi\rangle =-\bbox a\cdot\bbox b
\end{eqnarray}
which is the
nonrelativistic result. This case will never occur in a
realistic experiment since localization of detectors will lead
to a momentum spread. If $\bbox a\cdot\bbox n\neq 0$,
$\bbox b\cdot\bbox n\neq 0$ then in the ultrarelativistic case
$\beta^2=1$ 
\begin{eqnarray}
\langle \psi|\hat {\bbox a}\otimes \hat {\bbox
b}|\psi\rangle =-\frac{(\bbox a\cdot\bbox n)\,(\bbox b\cdot\bbox n)}
{|\bbox a\cdot\bbox n|\,|\bbox b\cdot\bbox n|}=\pm 1
\end{eqnarray}
independently of the choice of $\bbox a$, $\bbox b$. 

It is easy to intuitively understand this result: In the
ultrarelativistic limit projections of spin in directions
perpendicular to the momentum vanish for both particles and
spins are (anti)parallel to the momentum. 
The most striking case occurs if $\bbox a$ and $\bbox b$ are
perpendicular and the nonrelativistic average is 0. 
Let $\bbox a\cdot\bbox b=0$, $\bbox a\cdot\bbox n=
\bbox b\cdot\bbox n=1/\sqrt{2}$. 
Then
\begin{eqnarray}
\langle \psi|\hat {\bbox a}\otimes \hat {\bbox
b}|\psi\rangle =
-\frac{\beta^2}{2- \beta^2}.\label{19}
\end{eqnarray}
This average is 0 in the rest frame ($\beta=0$) and $-1$ for
$\beta =1$. Any observable deviation from 0 in an EPR-Bohm type
experiment would be an
indication that the operators $\bbox S$ and $\bbox Q$ are
physically correct observables and that massive spin-1/2
particles are extended in the sense that centers of mass and
charge do not coincide. 

\begin{figure}
\epsffile{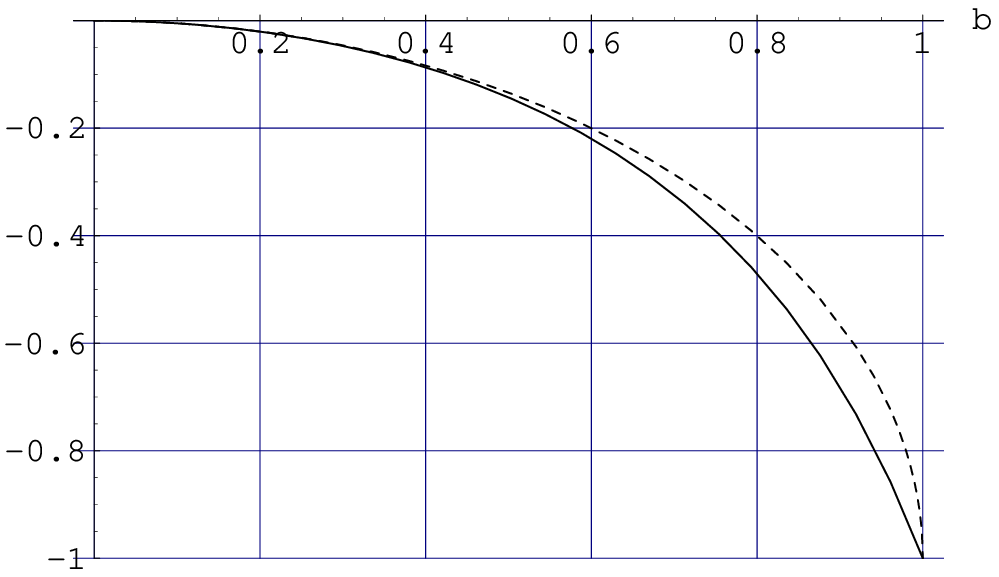}
\caption{
Average (18) (solid) as compared to $[1-\beta^2]^{1/2}-1$
(dotted). The EPRB average 
varies with $\beta$ faster than proper time. 
Relativistic corrections described by (15) and (18) are caused
by both the Lorentz contraction and the M{\o}ller shift of the
center of mass. Their experimental verification would provide an
indirect proof that the noncommuting position operator (9) is
physically well defined.
}
\label{fig3}
\end{figure}

Fig.~\ref{fig3} shows that (\ref{19}) describes a relativistic
effect that is even stronger than the Lorentz contraction or the time
delay (both are proportional to $\sqrt{1-\beta^2}$).

One pecularity of $\bbox Q$ is that its components
do not commute for nonzero spins. 
An uncertainty principle guarantees therefore that such a
particle cannot be localized at a point \cite{Kalnay}, or
is extended in some nonclassical sense,  
a property that cannot
be without implications for the renormalization and self-energy
problems. 
The definition of $\bbox
Q$ implies also that the center of mass does not transform as a
spatial component of a four-vector. This apparently counter-intuitive
result agrees however
with the classical analysis of M{\o}ller
\cite{Moller,Fleming1} who showed that the center of mass of a
spinning classical body is not a component of a four vector.
These interesting properties seem
unavoidable and can be proved in various ways at both quantum and classical
levels (for their classical derivations see \cite{Mukunda,Zakrzewski}).

\begin{figure}
\epsffile{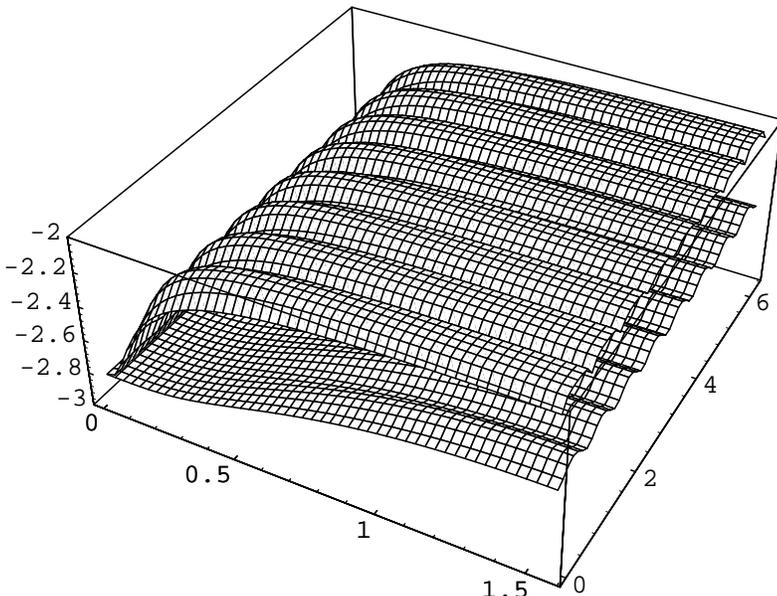}
\caption{The average (1) for $\beta=0.99$ (upper) and
$\beta=0.95$ (lower). $\theta=0$ corresponds to particles moving
perpendicularly to measuring devices (maximal violation).
For $\theta=\pi/2$ we have the situation from Fig.~2.}
\label{fig2}
\end{figure}

\begin{figure}
\epsffile{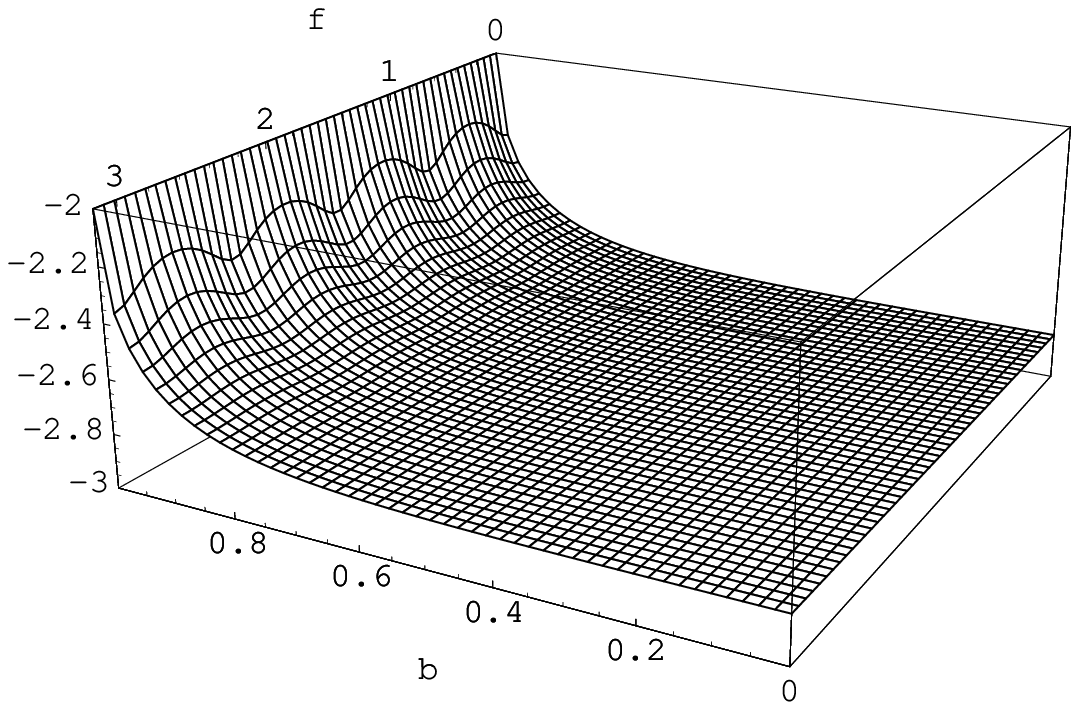}
\caption{For
$\beta=0$ we obtain the maximal violation and no violation
for $\beta\to 1$. Alice and Bob may be confused and ``detect" an
eavesdropper even if the state is pure singlet, but $\beta$ is
close to 1. 
Spins of ultrarelativistic spin-1/2 particles are
``almost classical" and are either almost parallel or anti-parallel to
their momenta}
\label{fig1}
\end{figure}

Consider now the vectors $\bbox a=(1/\sqrt{2},1/\sqrt{2},0)$,
$\bbox a'=(-1/\sqrt{2},1/\sqrt{2},0)$, $\bbox b=(0,1,0)$, 
$\bbox b'=(1,0,0)$ leading to the the maximal violation of the
Bell inequality in nonrelativistic domain. Fig.~\ref{fig1} shows the
dependence of the average (\ref{Bell}) on $\beta$ and
$\phi$ where $\bbox \beta=(\beta\cos\phi,\beta\sin\phi,0)$. 
Fig.~\ref{fig2} shows the average (\ref{Bell}) for $\bbox \beta=
\beta(\cos\phi\sin\theta,\sin\phi\sin\theta,\cos\theta)$ as a
function of the spherical angles and for $\beta=0.99$ and
$\beta=0.95$.

These results show clearly that the information about the degree
of violation of the Bell inequality is not sufficient for
determining purity of a massive two-particle zero-helicity state. 
Additionally one has
to know the momentum distribution of the particle beam.

\section{Pauli-Lubanski vector vs. spin}
\label{PL}

The relation between $W^\mu$ and $\bbox S$ is similar to the one
between the 4-velocity $u^\mu$ and the 3-velocity $\bbox \beta$.
The Casimir operator $W_\mu W^\mu$ equals $-(mc)^2j(j+1)$ if an
irreducible representation of the Poincar\'e group is
considered. For this reason it is typical to define the spin
4-vector as 
\begin{eqnarray}
w^\mu=W^\mu/(mc)=
\Bigl(\bbox u\cdot \bbox s, \frac{p_0}{mc}(\bbox n\cdot \bbox s)\bbox n
\pm  \bbox s_\perp\Bigr),
\end{eqnarray}
where $u^\mu$ is the 4-velocity.
In the rest frame $p_0=\pm mc$ and $w^\mu=
\pm(0,\bbox s)$ which seems to
justify this choice. For a moving particle the eigenvalues of
$\bbox a\cdot\bbox w$ are $\lambda_a\, p_0/mc$ where 
$\lambda_a$ denote the respective eigenvalues of 
$\bbox a\cdot\bbox S$. The eigenvalues of $\bbox a\cdot\bbox w$
therefore tend to $\pm\infty$ in the infinite momentum limit which
is unphysical for a spin observable. Nothing of
that kind occurs if one divides $\bbox W$ by {\it energy\/} and
{\it not by mass\/} which again selects our spin operator as a
candidate for a physical observable. 

Nevertheless, irrespective of this subtlety, the relativistic
EPRB average is the same for both $\bbox S$ and $\bbox w$ since
we consider a ``yes-no" observable which is obtained by {\it
normalization\/} of eigenvalues to $\pm 1$. This is another
reason to believe that the discussed suppression of degree of 
the Bell inequality
violation is a physical phenomenon that should be observable in
experiments with massive particles. 

\section{Comparison with the Dirac equation}

Just for the sake of completeness let us compare the general
formulas to the analogous calculations performed for the Dirac
electrons \cite{1984}. The Pauli-Lubanski vector is
\begin{eqnarray}
W^0&=&\bbox p\cdot \bbox s\\
\bbox W&=&\frac{1}{2}\bigl(\bbox s H + H \bbox s\bigr),
\end{eqnarray}
where $H$ is the Dirac free Hamiltonian and 
$$
\bbox s=\frac{\hbar}{2}
\left(
\begin{array}{cc}
\bbox \sigma & 0\\
0 & \bbox \sigma
\end{array}
\right)
$$
is the spinor part of the generator of rotations. The
relativistic spin operator is therefore equal to
\begin{eqnarray}
\bbox S&=&\bbox  W H^{-1}=
\frac{1}{2}\bigl(\bbox s + \Lambda \bbox
s\Lambda\bigr)\nonumber\\ 
&=&\Pi_+\bbox s\Pi_++ \Pi_-\bbox s\Pi_-.
\end{eqnarray}
Here $\Lambda$ is the sign-of-energy operator and $\Pi_\pm$
project on states of given signs of energy. 
It follows that $\bbox S$ is
the so-called even part of the spinor part of the generator of
rotations. This operator commutes with $H$ and hence can be used
for analyzing the EPRB experiment \cite{wrong}. 
The explicit form of this
observable in units with $\hbar=1$ and $c=1$ is
\begin{eqnarray}
\bbox S=
\frac{m^2}{p^2_0}\bbox s 
+
\frac{\bbox p^2}{p^2_0}(\bbox n\cdot \bbox s)\bbox  n
+
\frac{im}{2p^2_0}\bbox p\times \bbox \gamma.
\end{eqnarray}
$\bbox \gamma=(\gamma^1,\gamma^2,\gamma^3)$ where $\gamma^k$ 
are Dirac matrices.
The eigenvalues of $\bbox a\cdot \bbox S$ are
given by (\ref{aS}) with $j_3=\pm 1/2$. The corresponding
positive-energy eigenstates in the standard representation are 
$$
\Psi^a_\pm=
\left(
\begin{array}{c}
\sqrt{|p_0|+m}\biggl( (|\lambda_a|+\frac{1}{2}\bbox a\cdot \bbox n)
w_\pm \pm \frac{m a\cdot t}{2|p_0|}w_\mp\biggr)\\
\sqrt{|p_0|-m}\biggl(\pm (|\lambda_a|+\frac{1}{2}\bbox a\cdot\bbox n)
w_\pm -\frac{m a\cdot t}{2|p_0|}w_\mp\biggr)
\end{array}
\right)
$$
where $w_\pm$ satisfies $\bbox n\cdot\bbox \sigma w_\pm =\pm
w_\pm$, and $\bbox n\cdot\bbox t=0$. 

I have remarked that a positive verification of the relativistic
center-of-mass concept would indicate that nonzero spin
relativistic particles are extended.
The example of the Dirac equation illustrates this idea. 
Consider again the spinor part of the generator of rotations
$\bbox s$. It does not commute with $H$ and satisfies in the
Heisenberg picture the precession equation
\begin{eqnarray}
\dot {\bbox s}=\bbox \omega\times \bbox s,\label{prec}
\end{eqnarray}
where $\bbox \omega=-2c\gamma^5\bbox p/\hbar$. For massive
fields $\bbox \omega$ does not commute with $H$ and hence can be
decomposed into even and odd parts. The even part is
$$
\bbox{\Omega}={c^2+m\,c^3\bbox{\gamma}\cdot\bbox{n}/|\bbox{p}|\over
c^2+m^2c^4/\bbox{p}\,^2}\,\bbox{\omega}.
$$
$\bbox \Omega$ reduces to $\bbox \omega$ in both massless and
infinite-momentum limits. 
A Hamiltonian of a particle moving with velocity $\bbox{
v}=c\bbox{\beta}$ can now be expressed as
\begin{equation}
H=\Bigl(1+{m^2c^4\over c^2\bbox{p}^2}\Bigr) \bbox{\Omega}\cdot\bbox{S}
= \bbox{\beta}^{-2}\bbox{\Omega}\cdot\bbox{S}
= \bbox{\Omega'}\cdot\bbox{
S}\label{NH}
\end{equation}
where each of the operators appearing in $H$ is even and
commuting with $H$. 
The form (\ref{NH}) is analogous to the one discussed
in \cite{Hes}.
The limiting form $H=\bbox{\omega}\cdot\bbox{S}$
is characteristic of all massless fields, where for higher
spins the equation (\ref{prec}) is still valid, but angular velocities
for a given momentum are smaller the greater the
helicity.
The new form of the Hamiltonian leads to the following  
observation. Notice that for massless fields the 
Hamiltonian can be written in either of the following two forms
\begin{equation}
H=\bbox{\omega}\cdot\bbox{S}\label{H1}
\end{equation}
or
\begin{equation}
H=\bbox{c}\cdot\bbox{p}=\bbox{v}\cdot\bbox{p}\label{H2}
\end{equation}
where $\bbox{v}$ is the velocity operator for a general massless field
 ($c\bbox{\alpha}$ in case of the 
Dirac equation) and $\bbox{c}=(\bbox{v}\cdot\bbox{p})\bbox{p}/\bbox{p}\,^2$ is
its even part. We recognize here the classical mechanical rule for
a transition from a point-like description to the extended-object-like 
one: linear momentum goes into angular momentum, linear velocity into
angular velocity, and vice versa. The third part of this rule (mass--moment
of inertia) can be naturally postulated as follows
\begin{equation}
H=m_k\bbox{c}\,^2=I_k\bbox{\omega}\,^2.\label{H3}
\end{equation}
where (\ref{H3}) defines the kinetic mass $(m_k)$ and the kinetic moment 
of inertia $(I_k)$ of the massless field. The explicit form 
of $I_k$ for massless fields of helicity $\lambda=m-n$ [corresponding to
the $(m,n)$ spinor representation of $SL(2,C)$] is, in ordinary units,
\begin{equation}
I_k=\frac{\lambda\hbar\bbox{p}\cdot\bbox{S}}{c \bbox{p}\,^2}.\label{in}
\end{equation}
The  equation 
\begin{equation}
I_k=m_k r_\lambda^2
\end{equation}
characteristic, by the way, of circular strings (here with mass $m_k$) 
defines some radius which is equal to 
\begin{equation}
r_\lambda=\frac{\hbar \lambda}{|\bbox{p}|}\label{rad}
\end{equation}
which can be expressed also as a form of the 
uncertainty principle
\begin{equation}
{|\bbox{p}|} r_\lambda={\hbar \lambda}.\label{rad'}
\end{equation}
The center-of-mass commutation relation
$$
[Q_k,Q_l]=-i\hbar\epsilon_{klm}S_m/p_0^2
$$ 
leads in the massless
case to the
uncertainty relations of the type 
$$
\Delta Q_1\Delta Q_2\geq
\hbar^2|\lambda|/(2 \langle|\bbox p|^2\rangle|)=\langle
r_\lambda^2\rangle/|2\lambda|.
$$
It is  remarkable  that the same radius occurs naturally   in the 
 twistor formulation of massless fields \cite{PR}. 
It is known that although spin-0 twistors can be represented geometrically
by null straight lines, this does not hold for  spin-$\lambda$, 
$\lambda\neq 0$,
twistors \cite{PR}. Instead of the straight line we get a congruence 
of twisting, null, shear-free world lines, the so-called Robinson congruence.
A three-dimensional projection of this congruence consits of {\it circles\/},
whose radii are given exactly by our formula (\ref{rad}) (cf.~the footnote 
at p.~62 in  \cite{PR}). The circles propagate with velocity of light in the 
momentum direction and rotate in the right- or left-handed sense depending on
the sign of helicity. The same construction can be performed for
the massive Dirac particle if one uses $\bbox \Omega'$. 

\section{Summary}

The main idea advocated in this paper can be summarized as
follows. Consider {\it some\/} procedure leading to a
measurement of a nonrelativistic spin \cite{some}. This procedure is based
on a black box giving results ``yes" or ``no" for spins equal to,
respectively, $+\hbar/2$ and $-\hbar/2$. In the nonrelativistic
domain the particles enter the device ``slowly". Imagine now
that for some reasons we decide to use faster particles. 
The measured average may vary with the growing (average) 
velocity of the particle beam and, obviously, some result will
be obtained. The question is how to calculate the result of such
an experiment assuming that the
procedure measures the spin itself and not the total angular
momentum. Many different definitions of relativistic spin exist
in literature but all of them are momentum dependent
\cite{Bagrov}. Calculations based on 
the definition which seems the most physical
({\it via\/} the relativistic center-of-mass) show that
relativistic corrections are nontrivial. Their strength can be
regarded as a combined influence of two independent relativistic
phenomena: The Lorentz contraction and the M{\o}ller shift of
the center of mass of a spinning body.
The same result is
obtained if one uses the spin operator defined {\it via\/} the
Pauli-Lubanski vector. 
The effect can be in principle measured 
and will have to be taken into account in quantum cryptographic
tests for eavesdropping if fast massive particles will be used
for a key transfer.

\acknowledgements

I am grateful to Ryszard Horodecki for suggesting the problem,
Vasant Natarajan for informations concerning experiments, 
and Gerald Kaiser for extensive discussions. 
The paper is a part of the KBN project 2P30B03809.

\noindent

\begin{references}
\bibitem{wrong}The equivalence between 
the relativistic spin operator and the
even part of the spinor part of the generator of rotations may
lead to the naive conclusion that a measurable EPRB average could
be simply
$
\langle \psi| \tilde a\otimes 
\tilde b|\psi\rangle,
$
where $\tilde a=\vec a\cdot \vec s/(\hbar/2)$ etc.,
provided the electromagnetic fields used during the measurement
are sufficiently weak so that do not couple particle to
anti-particle states, and $\psi$ is a positive-energy
state\cite{second}.  
This assertion is wrong since 
the average uses the incorrectly normalized ``yes-no" observable:
The correct normalization is $\vec a\cdot \vec s/|\lambda_a|$
that is by eigenvalues of the relativistic operator.
\bibitem{second}I should mention here another argument appearing
sometimes in similar contexts. It is claimed that the second
quantization eliminates problems with negative energies so
no decomposition of operators into ``even" and ``odd" parts is
really physical. Notice, however, that we have obtained the
even spin of the Dirac particle 
as a by-product of the analysis which started from
unitary representations of the Poincar\'e group.  These are
precisely the representations that are used in quantum field
theory and the notion of the Pauli-Lubanski vector ``survives"
second quantization. The same concerns the
relativistic spin introduced above.
\end{references}
\end{document}